\documentclass[11pt,twoside]{article}
\usepackage{graphicx}
\usepackage{subfigure}
\usepackage{fancyheadings}
\usepackage{amsmath}
\usepackage{amssymb}
\usepackage{cite}
\usepackage{color,colortbl}

\definecolor{darkishgreen}{RGB}{39,203,22}
\definecolor{LightCyan}{rgb}{0.88,1,1}
\definecolor{Gray}{gray}{0.9}
\definecolor{lightRed}{RGB}{230,170,150}
\definecolor{modRed}{RGB}{230,82,90}
\definecolor{strongRed}{RGB}{230,6,6}

\begin{document}
\newcommand{\pst}{\hspace*{1.5em}}

\newcommand{\rigmark}{\em Journal of Russian Laser Research}
\newcommand{\lemark}{\em Volume 30, Number 5, 2009}

\newcommand{\be}{\begin{equation}}
\newcommand{\ee}{\end{equation}}
\newcommand{\bm}{\boldmath}
\newcommand{\ds}{\displaystyle}
\newcommand{\bea}{\begin{eqnarray}}
\newcommand{\eea}{\end{eqnarray}}
\newcommand{\ba}{\begin{array}}
\newcommand{\ea}{\end{array}}
\newcommand{\arcsinh}{\mathop{\rm arcsinh}\nolimits}
\newcommand{\arctanh}{\mathop{\rm arctanh}\nolimits}
\newcommand{\bc}{\begin{center}}
\newcommand{\ec}{\end{center}}

\thispagestyle{plain}

\label{sh}


\begin{center} {\Large \bf
\begin{tabular}{c}
Steering and correlations for the single qudit state 
\\[-1mm]
on the example of $j=3/2$
\end{tabular}
 } \end{center}

\bigskip

\bigskip

\begin{center} {\bf
V.I. Man'ko$^{1}$ and L.A. Markovich$^{2*}$
}\end{center}

\medskip

\begin{center}
{\it
$^1$P.N. Lebedev Physical Institute, Russian Academy of Sciences\\
Leninskii Prospect 53, Moscow 119991, Russia

\smallskip

$^2$Institute of Control Sciences, Russian Academy of Sciences\\
Profsoyuznaya 65, Moscow 117997, Russia
}
\smallskip

$^*$Corresponding author e-mail:~~~kimo1~@~mail.ru\\
\end{center}

\begin{abstract}\noindent
The phenomenon of quantum steering  and probabilistic meaning of the correlations are discussed for the state of the single qudit. The method of qubit
portrait of the qudit states is used to extend the known steering detection inequality to the system without subsystems. The example of the $X$-state with $j=3/2$ is studied in detail.\end{abstract}

\medskip

\section{Introduction}
\pst
The notion of the quantum steering was introduced by E. Schr\"{o}dinger \cite{Schrodinger} as an answer to the famous paper of A. Einstein, B. Podolsky and N. Rosen \cite{Einstein} to generalize the EPR paradox. A good overview on these papers is given in \cite{Werner2}. Although, the problem was posed in 1935, the concept of the quantum steering remains a popular topic. It seem to be a rich resource in numerous application like one-way quantum cryptography \cite{Chen,Saunders} or visualization of the two-qubit state tomography \cite{Jevtic}.  In \cite{Marciniak} the EPR steering was considered as a form of nonlocality in the quantum mechanics, that is something in between the entanglement and the Bell nonlocality. Interesting to say that not all entangled states are steered, and not every steerable state violate the Bell inequality. The EPR steering can be detected through the violation of the steering inequalities \cite{Saunders,Zukowski,Schneeloch,Schneeloch2}. However, in \cite{Chen} it was shown that there are many quantum steerable states which do not violate the known steering inequalities.
\par The notion of steering is highly connected with the notion of the correlation function. In \cite{Zukowski} the steering inequality is based on the maximum of the correlation function. In \cite{Wiseman} the steering concept was reformulated. The steering was considered as the ability of the first system to affect the state of the second system through the choice of the first systems measurement basis.  Since that, the concept of the quantum steering can be introduced not only for the multipartite (joint) systems, but for all those systems with correlations.
\par Recently, it was observed in \cite{Chernega,Chernega14,OlgaMankoarxiv} that the quantum properties of the systems without subsystems (single qudit) can be formulated by using an invertible map of integers $1,2,3\ldots$ onto the pairs (triples, etc) of integers $(i,k)$, $i,k=1,2,\ldots$ (or semiintegers). For example, the single qudit state $j=0,1/2,1,3/2,2,\ldots$ can be mapped onto the density operator of the system containing the subsystems like the state of the two qubits. Using this mapping, the notion of the separability and the entanglement was extended in \cite{Markovich3} to the case of the single qudit $X$-state with $j=3/2$. With the help of the latter mapping we can introduce the analog of the correlation function for the single qudit system. Hence, the notion of the steering for the system without subsystems can be obtained.
\par The paper is organized as follows. In  Sec. \ref{sec:1} the notion of the EPR steering is extended to the case of the system without subsystems (single qudit). One of the known steering inequalities \cite{Zukowski} is generalized to the case of the single qudit with the spin $j=3/2$. In  Sec. \ref{sec:2} the probabilistic meaning of the correlations in the single qudit system is given. In  Sec. \ref{sec:3} the latter results are illustrated on the examples of the two $X$-states - the Werner and the Gisin states of the single qudit with the spin $j=3/2$.
\section{Quantum Steering}\label{sec:1}
Let us define the quantum state described by the following $4\times4$ density matrix
\begin{eqnarray}\rho=\left(
                                 \begin{array}{cccc}
                                   \rho_{11}& \rho_{12}& \rho_{13}& \rho_{14}\\
                                   \rho_{21}& \rho_{22}& \rho_{23}& \rho_{24}\\
                                   \rho_{31}& \rho_{32}& \rho_{33}& \rho_{34}\\
                                   \rho_{41}& \rho_{42}& \rho_{43}& \rho_{44}\\
                                 \end{array}
                               \right). \label{1}
                               \end{eqnarray}
The matrix (\ref{1}) has the standard properties, namely, $\rho=\rho^{\dagger}$, $Tr\rho=1$ and its eigenvalues are nonnegative. The diagonal
elements of the density matrix (\ref{1}) can be considered as the components of the probability vector $\overrightarrow{p}=(p_{11},p_{22},p_{33},p_{44})$, $\sum\limits_{i}p_{ii}=1$, $0\leq p_{ii}\leq1$. If we use the invertible mapping of indices $1\leftrightarrow 1/2~1/2$; $2\leftrightarrow1/2~-1/2$; $3\leftrightarrow-1/2~1/2$; $4\leftrightarrow-1/2~-1/2$, then the matrix (\ref{1}) can describe the two-qubit state. In the case of the two particles the correlations in the system can be described by the joint probability function
 \begin{eqnarray*}P(a,b|A, B)=\int p_{\lambda}P(a|A,\lambda)P(b|B,\lambda)d\lambda,
     \end{eqnarray*}
     where $P(a|A,\lambda)$ is a probability distribution of  the measurement outcomes $a$ under setting $A$ for a hidden variable $\lambda$. The hidden variable has the probability distribution $p_{\lambda}$ and its hidden state is $\rho_{\lambda}$. If the following model of the correlation do not exist
      \begin{eqnarray*}P(a,b|A, B)=\!\int p_{\lambda}P(a|A,\lambda)Tr(\widehat{\pi}(b|B)\rho^{(b)}_{\lambda})d\lambda,
     \end{eqnarray*}
     then the state is steerable \cite{Zukowski}. The  $\widehat{\pi}(b|B)$ is the projection operator for an observable parameterized by the setting $B$ and the $\rho^{(b)}_{\lambda}$ is some pure state of the system $B$.
     \par The EPR steering can be detected through the violation of the steering inequalities. In \cite{Zukowski} the steering inequality is based on the maxima of the correlation function. The quantum correlation function for the two-qubit state is given by
 \begin{eqnarray} \label{4}E(\overrightarrow{m},\overrightarrow{n})&=&Tr(\overrightarrow{m}\cdot\overrightarrow{\sigma}\otimes\overrightarrow{n}\cdot\overrightarrow{\sigma}\rho)\nonumber\\
 &=&\sum\limits_{i,j=1}^{3}T_{ij}m_in_j,
    \end{eqnarray}
 where  $\overrightarrow{\sigma}$ is the vector built out of the Pauli matrix, $\overrightarrow{m}=(m1,m2,m3)$, $\overrightarrow{n}=(n1,n2,n3)$ are the unit Bloch vectors, $T_{ij}$ are the components of the correlation matrix. It is known, that if the state  is non-steerable, then the following inequality is fulfilled
   \begin{eqnarray}\label{5}\max_{\overrightarrow{m},\overrightarrow{n}}(E(\overrightarrow{m},\overrightarrow{n}))&\geq&\frac{2}{3}\sum\limits_{i,j=1}^{3}T_{ij}.
       \end{eqnarray}
       Let us extend the latter steering inequality to the case of the system without subsystems.
 \par Applying the invertible map of indices $1\leftrightarrow 3/2$, $2\leftrightarrow1/2$, $3\leftrightarrow-1/2$, $4\leftrightarrow-3/2$, the density matrix (\ref{1}) can be rewritten as
 \begin{eqnarray}\rho_{\frac{3}{2}}=\left(
                                 \begin{array}{cccc}
                                   \rho_{\frac{3}{2},\frac{3}{2}}& \rho_{\frac{3}{2},\frac{1}{2}}& \rho_{\frac{3}{2},-\frac{1}{2}}& \rho_{\frac{3}{2},-\frac{3}{2}}\\
                                   \rho_{\frac{1}{2},\frac{3}{2}}& \rho_{\frac{1}{2},\frac{1}{2}}& \rho_{\frac{1}{2},-\frac{1}{2}}& \rho_{\frac{1}{2},-\frac{3}{2}}\\
                                   \rho_{-\frac{1}{2},\frac{3}{2}}& \rho_{-\frac{1}{2},\frac{1}{2}}& \rho_{-\frac{1}{2},-\frac{1}{2}}& \rho_{-\frac{1}{2},-\frac{3}{2}}\\
                                   \rho_{-\frac{3}{2},\frac{3}{2}}& \rho_{-\frac{3}{2},\frac{1}{2}}& \rho_{-\frac{3}{2},-\frac{1}{2}}& \rho_{-\frac{3}{2},-\frac{3}{2}}\\
                                 \end{array}
                               \right)\label{6}
                               \end{eqnarray}
and can describe the noncomposite system of the single qudit with the spin $j=3/2$.
The matrix saves the standard properties of the density matrix, \emph{e.g.} $\rho_{3/2}=\rho_{3/2}^{\dagger}$, $Tr\rho_{3/2}=1$ and its eigenvalues are nonnegative.
 It means that all the equalities and inequalities known for the matrix (\ref{1} (the two-qubit system) are valid for the matrix (\ref{6}) (the single qudit system). Hence, the quantum correlation function (\ref{4}) can be written also for the single qudit state
 and its correlation tensor has the form
 \begin{eqnarray}T_{11}&=&\rho_{14} + \rho_{23} + \rho_{32} + \rho_{41},\nonumber\\
 T_{12}&=&(\rho_{14} - \rho_{23} + \rho_{32} - \rho_{41})i,\nonumber\\
 T_{13}&=&\rho_{13} - \rho_{24} + \rho_{31} - \rho_{42},\nonumber\\
  T_{21}&=&(\rho_{14} + \rho_{23} - \rho_{32} - \rho_{41})i,\nonumber\\
   T_{22}&=&\rho_{23} - \rho_{14} + \rho_{32} - \rho_{41},\\
    T_{23}&=&(\rho_{13} - \rho_{24} - \rho_{31} + \rho_{42})i,\nonumber\\
     T_{31}&=&\rho_{12} + \rho_{21} - \rho_{34} - \rho_{43},\nonumber\\
      T_{32}&=&(\rho_{12} - \rho_{21} - \rho_{34} + \rho_{43})i,\nonumber\\
       T_{33}&=&\rho_{11} - \rho_{22} - \rho_{33} + \rho_{44}.\nonumber\label{2}
 \end{eqnarray}
 Needless to say that the meaning of the correlation function for the single qudit state differs from  the standard correlation function of the bipartite states.
 \section{Correlations in the Single Qudit State}\label{sec:2}
 Let us first start from the classical example of the two coins which can drop on the one ($1$) or on the second ($2$) side. Hence we have two random variables $m_1$, $m_2$  and four opportunities $(m_1,m_2)=\{(11), (12), (21), (22)\}$ with the probabilities $p_{ij},\quad i,j=1,2$. Let $\omega(m_1,m_2)$ be the joint probability function of these two random variables. There marginal probability functions can be defined as
 \begin{eqnarray*}\omega_1(m_1)&=&\sum\limits_{m_2}\omega(m_1,m_2),\\
  \omega_2(m_2)&=&\sum\limits_{m_1}\omega(m_1,m_2).\end{eqnarray*}
 Hence, the correlation between the two observations is given by $\langle m_1,m_2\rangle=\sum\limits_{m_1,m_2}m_1m_2\omega(m_1,m_2)$.
 \par However, we can be interested only in the cases when the first coin falls on the one side and the second coin is not interesting for us. We have the two new outcomes $\{\widetilde{\omega}\}$ with the probabilities  $\widetilde{p}_1=p_{11}+p_{12}$ and $\widetilde{p}_2=p_{22}+p_{21}$. If we are interested only in the second coin, analogously we have the two new outcomes $\omega$ with the probabilities $p_1=p_{11}+p_{21}$ and $p_2=p_{22}+p_{12}$. The outcomes $\{\omega\}$ and $\{\widetilde{\omega}\}$
  are correlated. The latter example is helpful to show the existence of the correlations in the systems without subsystems.
 \par If we have the single qudit system with the spin $j=3/2$, we can write the sample space $\Omega$ of the four outcomes $\omega\in\Omega$ (the values of the spin projection)
  $|m>=\{|3/2>, |1/2>, |-1/2>, |-3/2>\}$ with the probabilities $p_{3/2},p_{1/2},p_{-1/2},p_{-3/2}$. Then we have one four-level atom, \emph{i.e.} the
$|m>=|3/2>$  corresponds to the case when the highest (fourth) level is filled.
If we are interested only in the outcomes when the fourth or the second level of the four-level atom is filled, then we can assume that we have the new set of two outcomes $\{\omega_1\}$ with the probabilities $p_1=p_{3/2}+p_{-1/2}$,  $p_2=p_{-3/2}+p_{1/2}$. If we are interested only in the outcomes when the fourth and the third level is filled then we have another set of the outcomes $\{\omega_2\}$ and there probabilities are $\widetilde{p}_1=p_{3/2}+p_{1/2}$,  $\widetilde{p}_2=p_{-3/2}+p_{-1/2}$. The outcomes $\{\omega_1\}$ and $\{\omega_2\}$ are correlated. Hence, the correlations in the single qudit systems are between the different combinations of the outcomes.
 \section{The $X$-state for the single qudit}\label{sec:3}
The single qudit state with the spin $j=3/2$ can be described by the $X$-state density matrix.
 \begin{eqnarray}\label{7}
\rho^{X}&=&\left(
                     \begin{array}{cccc}
                       \rho_{11} & 0 & 0 & \rho_{14}\\
                       0 & \rho_{22}& \rho_{23} & 0 \\
                       0 & \rho_{23}^{\ast} & \rho_{33} & 0 \\
                      \rho_{14}^{\ast} & 0 & 0 & \rho_{44} \\
                     \end{array}
                   \right),
\end{eqnarray}
where $\rho_{11},\rho_{22},\rho_{33},\rho_{44}$ are positive reals and $\rho_{23},\rho_{14}$ are complex quantities.
The matrix (\ref{7}) has the unit trace and it is nonnegative if $\rho_{22}\rho_{33}\geq|\rho_{23}|^2$, $\rho_{11}\rho_{44}\geq|\rho_{14}|^2$ hold.
For the latter density matrix the correlation tensor (\ref{2}) changes
 \begin{eqnarray}
 T_{13}= T_{23}= T_{31}=T_{32}=0.\label{3}
 \end{eqnarray}
\par
\begin{figure}
\begin{center}
\includegraphics[width=8.4cm]{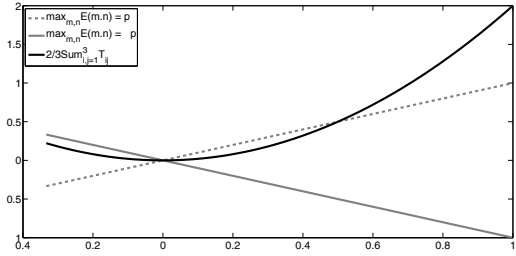}
\caption{The left (gray) and the right hand side (black) of the inequality (\ref{5}) for the Werner state} \label{fig:1}
\end{center}
\end{figure}

 One of the examples of the $X$-state is given by the Werner state \cite{Werner}. The single qudit with the spin $j=3/2$ can be described by the following Werner density matrix
\begin{eqnarray}\label{10_1}\rho^{W}&=&\left(
                     \begin{array}{cccc}
                       \frac{1+p}{4} & 0 & 0 & \frac{p}{2}\\
                       0 & \frac{1-p}{4}& 0 & 0 \\
                       0 & 0& \frac{1-p}{4} & 0 \\
                       \frac{p}{2} & 0 & 0 & \frac{1+p}{4} \\
                     \end{array}
                   \right),
\end{eqnarray}
where the parameter $p$ satisfies the inequality $-1/3\leq p\leq1$. The parameter domain $1/3< p\leq1$ corresponds to the entangled state.
The correlation tensor (\ref{2}) is the diagonal matrix with entries $p(1-1 \quad 1)$ and the maximum value of the correlation function (\ref{4} is $p$ in the domain $0<p<1/3$ and $-p$ in $-1/3<p\leq0$. Hence, the inequality (\ref{5}) is fulfilled if
$0<p<1/2$. Since we are interested only in the entangled states, the parameter domain $1/3<p<1/2$ corresponds to the steerable state. The left and the right hand side of the inequality (\ref{5}) are presented in fig.~\ref{fig:1}.
\par The second example is given by the single qudit state described by the Gisin density matrix
\begin{eqnarray}\label{10_2}\rho^{G}&=&\left(
                     \begin{array}{cccc}
                       \frac{1-x}{2} & 0 & 0 & 0\\
                       0 & x|a|^2& xab^{\ast} & 0 \\
                       0 & xa^{\ast}b& x|b|^2 & 0 \\
                       0 & 0 & 0 & \frac{1-x}{2} \\
                     \end{array}
                   \right),
\end{eqnarray}
where $|a|^2+|b|^2=1$, $0<x<1$. In \cite{Peres} it is shown that the latter matrix has positive eigenvalues if $x\leq(1+2|ab|)^{-1}$ holds. We denote it as $x_{max}=(1+2|ab|)^{-1}$. Let us select $a=0.2$. Then the correlation tensor (\ref{2}) is the diagonal matrix with entries $(4\sqrt{6}x/25\quad 4\sqrt{6}x/25 \quad 1 - 2x)$
and $x_{max}=0.718$.
\begin{figure}
\begin{center}
\includegraphics[width=8.4cm]{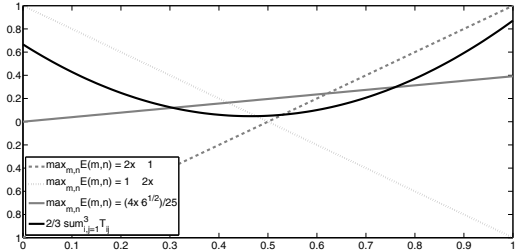}
\caption{The left (gray) and the right hand side (black) of the inequality (\ref{5}) for the Gisin state} \label{fig:2}
\end{center}
\end{figure}
The maximum value of the correlation function (\ref{4}) reads
\begin{eqnarray}&&2x - 1,\quad x>0.622;\quad
1 - 2x,\quad x<0.418;\nonumber\\
&&4\sqrt{6}x/25,\quad 0.418<x<0.622.
\end{eqnarray}
Hence, the inequality (\ref{5}) is fulfilled for all values of the parameter $x<x_{max}$. The left and the right hand side of the inequality (\ref{5}) for the Gisin state are presented in fig.~\ref{fig:2}.

 \section{Summary}
 To conclude we point out the main results of the work. We have applied the invertible mapping techniques \cite{Chernega} to introduce the steering notion for the systems without subsystems. We have shown that there exists the correlation function for such systems. Certainly, the probabilistic meaning of the latter correlations is different from the known correlations between the subsystems in the many partite systems. In the single qudit systems the correlations are between the different combinations of the outcomes.  Using the latter correlation function the steering detection inequality known for the bipartite systems \cite{Zukowski} can be formulated for the systems without subsystems (single qudit). However, using the invertible mapping any of the known steering inequalities can be obtained for the systems without subsystems. The obtained results were illustrated by the classical examples of the Werner and the Gisin $X$-states. Therefore, we have shown that the concept of the quantum steering can be applied to any quantum system both with the subsystems and the single qudit.

\end{document}